\documentclass[aps,prc,preprint,showpacs]{revtex4}
\usepackage{graphicx}
\usepackage[dvips]{color}

\begin{document}

\title{Neutrino and antineutrino CCQE scattering
in the SuperScaling Approximation from MiniBooNE to NOMAD energies}
\author{J.E. Amaro}
\affiliation{Departamento de F\'{\i}sica At\'{o}mica, Molecular y Nuclear
and Instituto Carlos I de F\'{\i}sica Teorica y Computacional, Universidad de Granada,
  18071 Granada, SPAIN}
\author{M.B. Barbaro}
\affiliation{Dipartimento di Fisica, Universit\`a di Torino and
  INFN, Sezione di Torino, Via P. Giuria 1, 10125 Torino, ITALY}
\author{J.A. Caballero}
\affiliation{Departamento de F\'{\i}sica At\'{o}mica, Molecular y Nuclear,
Universidad de Sevilla,
  41080 Sevilla, SPAIN}
\author{G.D. Megias}
\affiliation{Departamento de F\'{\i}sica At\'{o}mica, Molecular y Nuclear,
Universidad de Sevilla,
  41080 Sevilla, SPAIN}
\author{T.W. Donnelly}
\affiliation{Center for Theoretical Physics, Laboratory for Nuclear
  Science and Department of Physics, Massachusetts Institute of Technology,
  Cambridge, MA 02139, USA}

\begin{abstract}
We compare the predictions of the SuperScaling model for charged
current quasielastic muonic neutrino and antineutrino scattering
from $^{12}$C with experimental data spanning an energy range 
up to 100 GeV. 
We discuss the sensitivity of the results to different
parametrizations of the nucleon 
vector and axial-vector
 form factors. Finally, we show the
differences between electron and muon (anti-)~neutrino cross
sections relevant for 
the $\nu$STORM facility.
\end{abstract}

%\date{ }


\maketitle

\section{Introduction}

Charged-current (CC) quasielastic (QE) muonic neutrino and
antineutrino cross sections on a $^{12}$C target have been recently
measured by the MiniBooNE collaboration at
Fermilab~\cite{AguilarArevalo:2010zc,AguilarArevalo:2013hm} for
neutrino energies in the 1 GeV region, the neutrino and antineutrino
fluxes peaking at 0.79 GeV and 0.66 GeV, respectively, and going
from 0 to about 7 GeV with the most important contributions coming
from below 3 GeV. The results have stimulated many theoretical
studies~\cite{Amaro:2010sd,Amaro:2011qb,Amaro:2011aa,
Benhar:2010nx,Martini:2010ex,Juszczak:2010ve,Butkevich:2010cr,Meucci:2011ce,
Meucci:2011vd,Bodek:2011ps,Nieves:2011yp,Lalakulich:2012ac,Sobczyk:2012ms}
that attempt to explain the discrepancy between the data and
traditional nuclear models; these include the Relativistic Fermi Gas
(RFG) model, RPA calculations, the use of realistic spectral
functions, relativistic Green's function approaches and relativistic
mean field theory.

An empirical solution to this puzzle, proposed by the MiniBooNE
collaboration, advocates a value of the nucleon axial-vector dipole
mass $M_A\simeq 1.35$ GeV/c$^2$~\cite{AguilarArevalo:2010zc}, which
is significantly larger than the standard value $M_A=$1.032
GeV/c$^2$. On the other hand microscopic explanations based on
multi-nucleon excitations, in particular two-particle emission, were
proposed in
\cite{Amaro:2010sd,Amaro:2011qb,Martini:2010ex,Nieves:2011yp}. Those
of \cite{Martini:2010ex,Nieves:2011yp}, although rather different in
their basic ingredients, have been shown to give very good agreement
with the MiniBooNE data, while those of
\cite{Amaro:2010sd,Amaro:2011qb}, which are based on the exact
relativistic evaluation of the Meson Exchange Currents (MEC) within
the 2p2h RFG approach, provide an enhancement of the cross sections
but do not fully account for the discrepancy. It should be stressed
that a consistent evaluation of the MEC contribution is technically
hard to achieve and an exact relativistic gauge invariant
calculation of both vector and axial-vector contributions to MEC in
neutrino scattering is not yet available.

On the other hand, CCQE $\nu_\mu-$ and $\bar\nu_\mu-^{12}$C cross
section measurements from the NOMAD
collaboration~\cite{Lyubushkin:2008pe} for higher beam energies,
going from 3 to 100 GeV, do not call for an anomalously large
axial-vector mass and do not appear to match with the lower-energy
MiniBooNE results, as shown in Fig.~1 where the two sets of data are
displayed. 
It should also be mentioned that recent data on CCQE 
$\nu_\mu-$ and $\bar\nu_\mu-^{12}$C from the MINER$\nu$A 
Collaboration~\cite{Fields:2013zhk,Fiorentini:2013ezn} are claimed to
disfavor the value $M_A\simeq 1.35$ GeV/c$^2$.
It is thus desirable to perform a consistent theoretical
analysis of the cross sections in the entire 0-100 GeV energy range,
using a nuclear model which can be applied up to very high energies.
Such a model must obviously fulfill two basic requirements: it has
to be relativistic and it must successfully describe QE electron
scattering data from intermediate up to very high energies.

The SuperScaling (SuSA) model, based on the superscaling function
extracted from QE electron scattering data, does a reasonable job of
satisfying both of the above requirements: it is fully relativistic
and has been constructed using those data as input. On the one hand,
its applicability may be questioned at very low energies (meaning by
that, momentum transfers $q\le$ 400 MeV/c and energy transfers
$\omega\le$ 50 MeV), where collective effects which violate scaling
dominate. In discussing the results found (see below) we shall
return to comment on this issue. On the other hand the SuSA approach
can be safely extended up to very high energies, since it is based
on $(e,e')$ data in a range going from intermediate to high energies
and momentum
transfers~\cite{Day:1990mf,Donnelly:1998xg,Donnelly:1999sw,Maieron:2001it}.

In summary, the model gives a very good representation of all
existing QE electron scattering data for high enough momentum and
energy transfers, to the extent that {\em quasielastic} scattering
can be isolated. Additionally, the same scaling approach has been
shown to be very successful when extended to higher energies into
the non-QE regime where {\em inelastic} contributions
dominate~\cite{Maieron:2009an}. However, it does not account for the
typically $10-20\%$ scaling violations that occur mainly in the
transverse channel and are associated with non-impulsive processes
induced by two-body meson-exchange currents (MEC)
(see~\cite{Donnelly:1978xa,Van:81,De Pace:2003xu,Amaro:2010iu}).
These should therefore be added to obtain a representation of all of
the contributions to the inclusive cross section in the relevant
kinematical regions. While these have been included in our past
studies of the differential neutrino cross sections, where one can
be sure that the momentum and energy transfers are sufficiently
large that the modeling is robust, in this paper we neglect such
contributions since, as discussed below, the low-$q$/low-$\omega$
region is important for the total cross section. We are not
confident of the validity of the present 2p2h MEC model in this kinematic region --- work is
in progress to correct this deficiency.

The SuSA model has been extensively described in previous work (see,
e.g., \cite{Amaro:2004bs}).  In this paper we only summarize the
basic ideas and focus on applying them to CCQE (anti)neutrino
scattering from $^{12}$C, comparing the results with the MiniBooNE
and NOMAD data. We also study the sensitivity of the cross section
to different up-to-date parametrizations of the nucleon form factors
entering the cross section, $G_E$, $G_M$ and $G_A$, studying in
particular the effects of a monopole parametrization for the
axial-vector form factor. Finally, we present the SuSA predictions
for electron neutrino and antineutrino cross sections, with
particular reference to the $\nu$STORM kinematical
conditions~\cite{Kyberd:2012iz}.

\begin{figure}[htcb]
\includegraphics[scale=0.41, bb=20 35 722 550, clip]{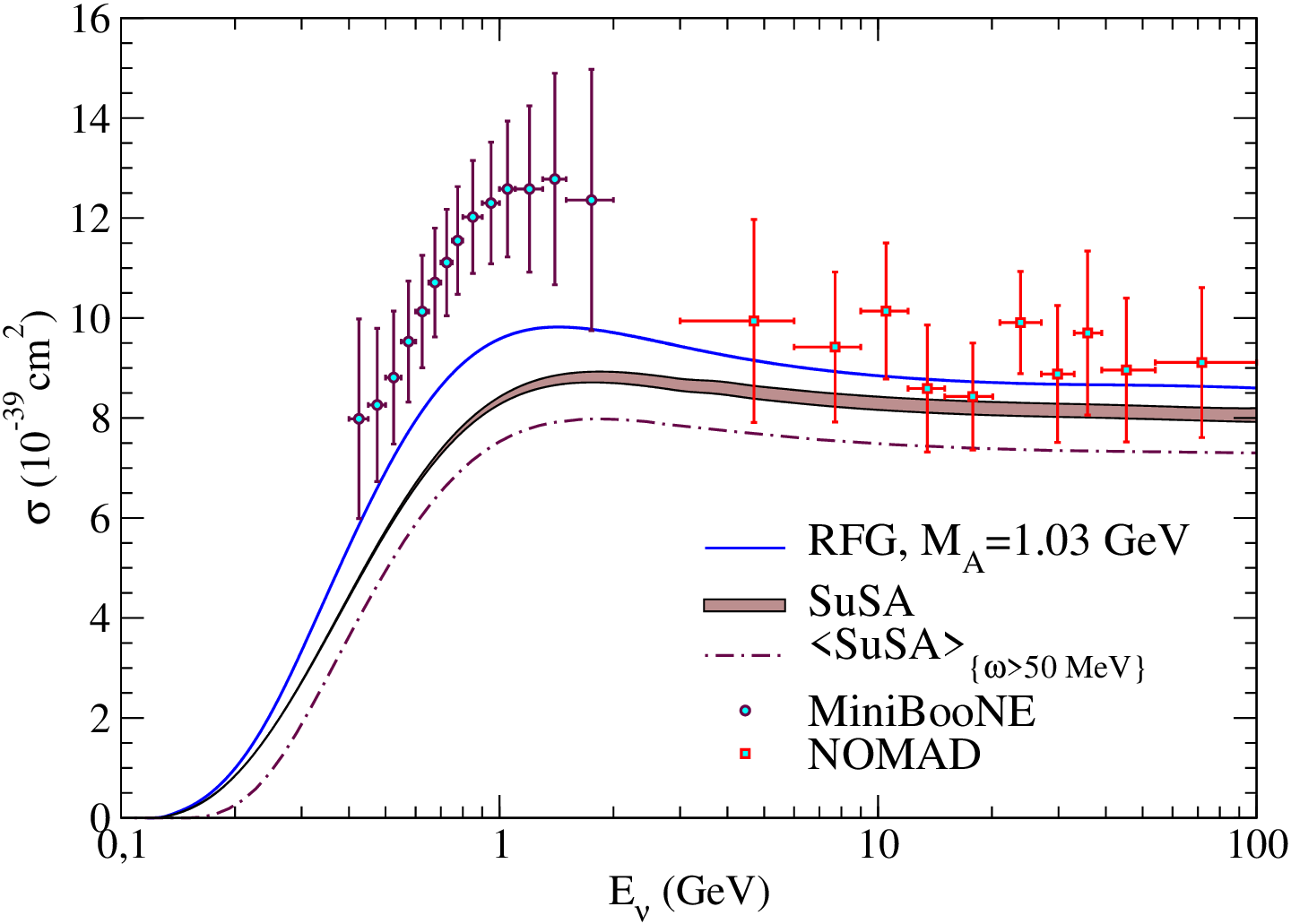}
\caption{(Color online) CCQE $\nu_\mu-^{12}$C  cross section per
nucleon displayed versus neutrino energy $E_\nu$ and evaluated using
the SuSA model (brown band) with the standard value of the
axial-vector dipole mass $M_A=1.032$ GeV/c$^2$. Results are also
shown for the RFG model with $M_A=1.032$ GeV/c$^2$ (blue solid curve) and compared
with the MiniBooNE~\cite{AguilarArevalo:2010zc} and
NOMAD~\cite{Lyubushkin:2008pe} experimental data. Also presented for reference are
the results for SuSA excluding all contributions coming from transferred energies below 50 MeV (dot-dashed curve).
\label{fig:sigmanumu}}
\end{figure}
\begin{figure}[htcb]
\includegraphics[scale=0.41, bb=20 35 722 550, clip]{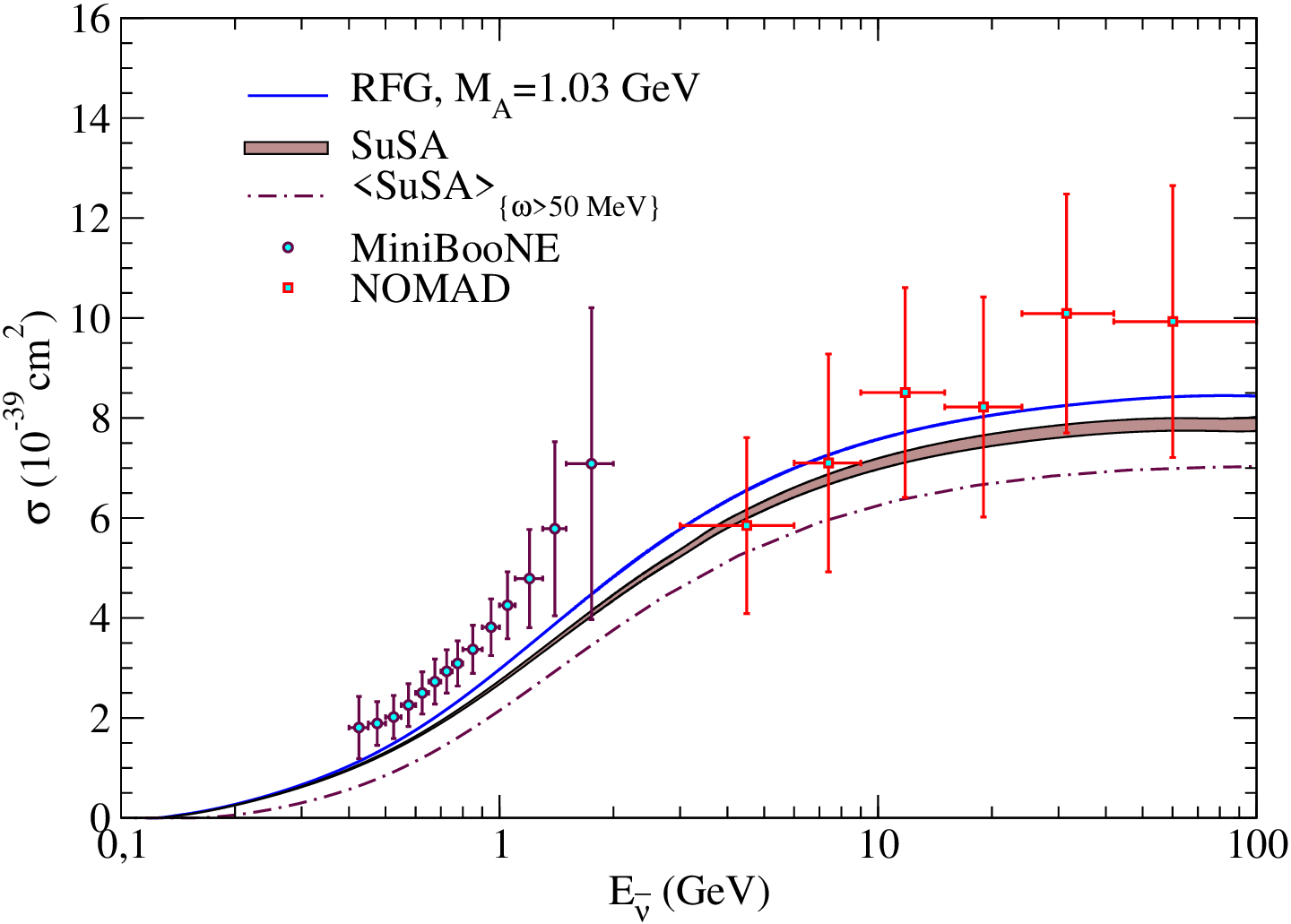}
\caption{(Color online) As for Fig.1, 
but now for $\overline\nu_\mu-^{12}$C scattering. The MiniBooNE data
are from \cite{AguilarArevalo:2013hm}.
\label{fig:sigmanumubar}}
\end{figure}
\begin{figure}[htcb]
\begin{center}
\includegraphics[scale=0.41, bb=20 35 722 550, clip]{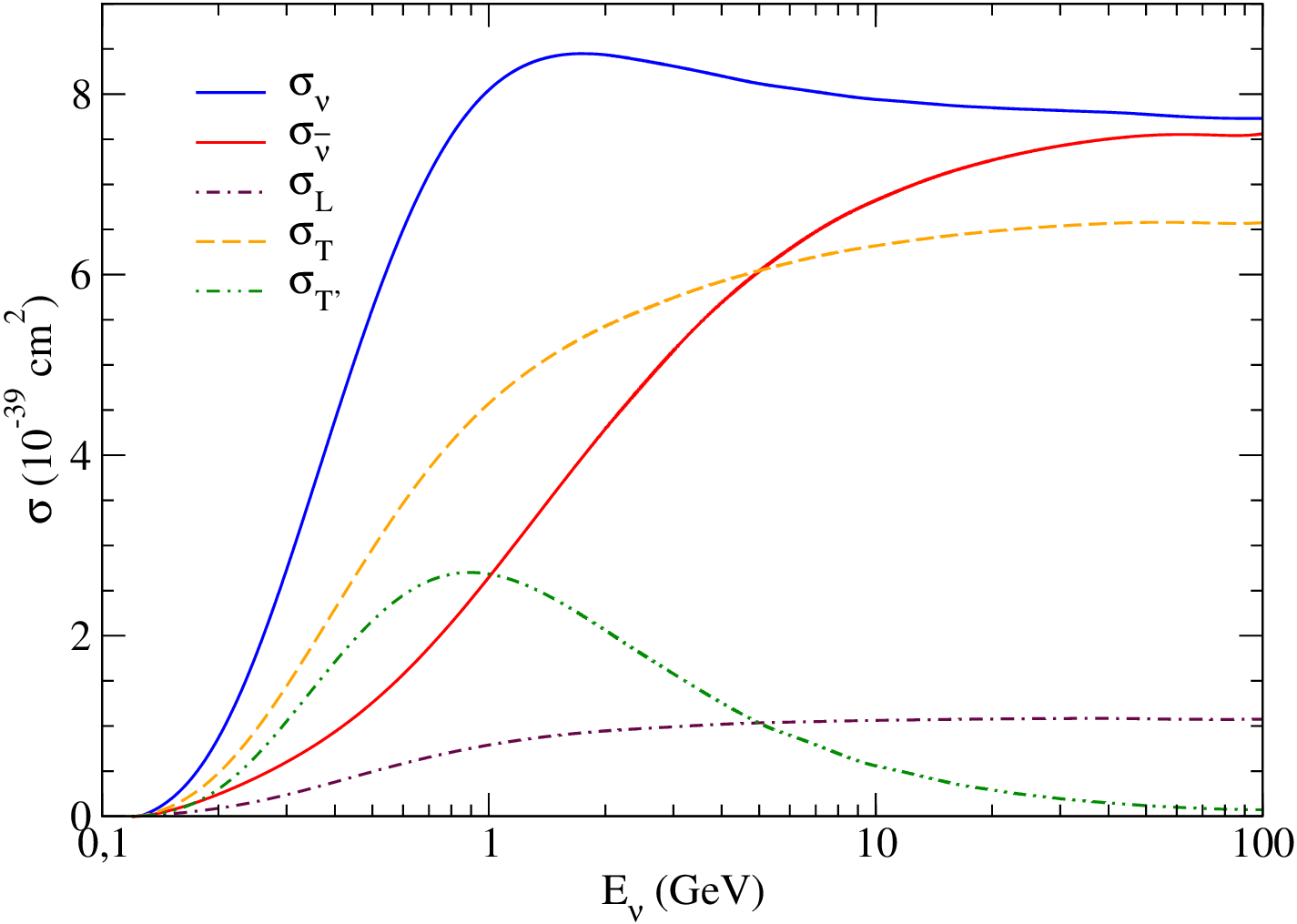}
\end{center}
\vspace{-0.4cm}
\caption{(Color online) Separated contributions in the SuSA model.
\label{fig:sep}}
\end{figure}
\begin{figure}[ht]
\includegraphics[scale=0.41, bb=20 0 722 635, clip]{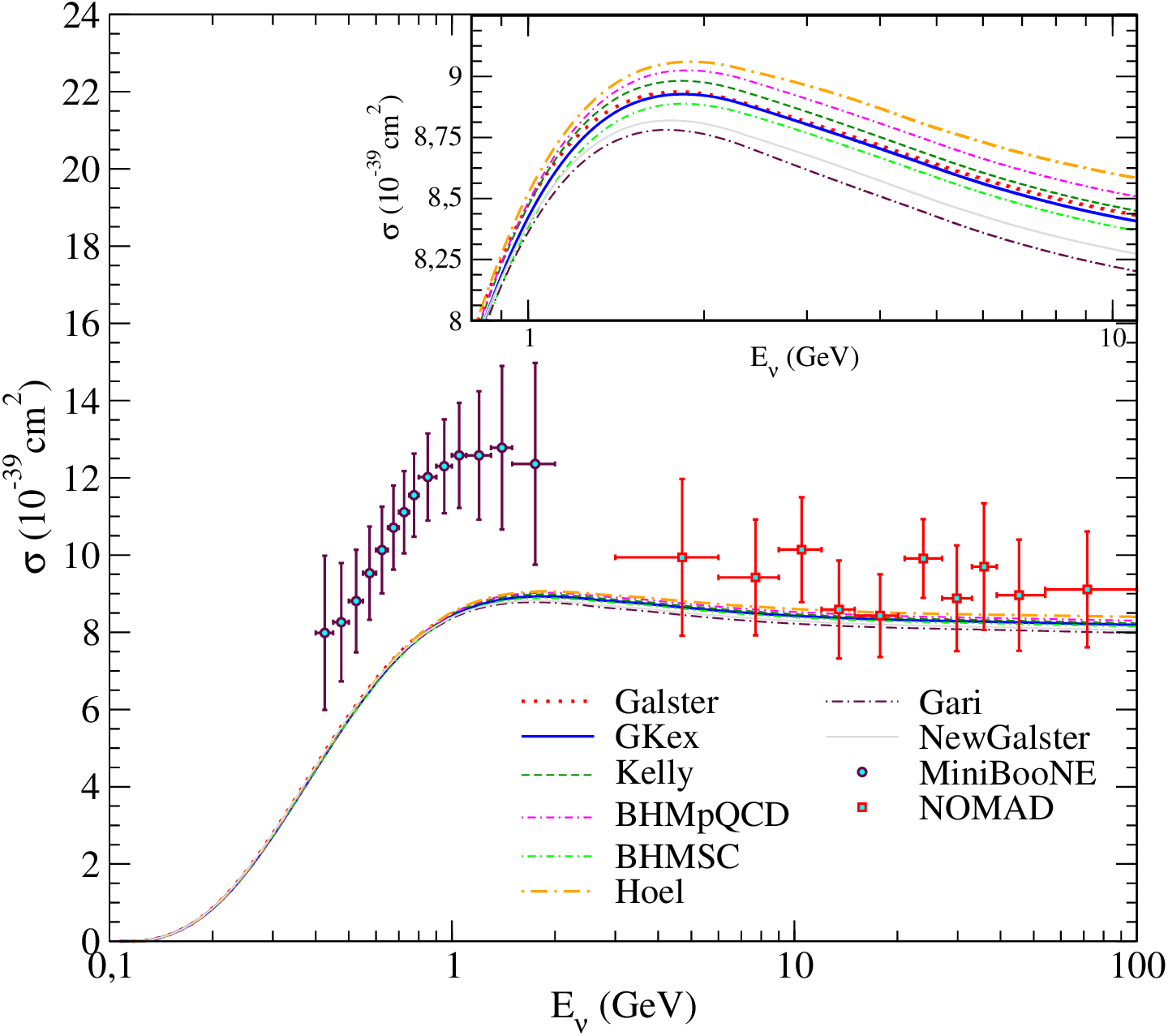} 
\caption{(Color online) CCQE $\nu_\mu-^{12}$C  cross section per
nucleon evaluated in the SuSA model for various parametrizations of
the nucleon electromagnetic form factors. 
\label{fig:emffnu}}
\end{figure}

\section{Results}

The SuSA approach to neutrino scattering is based on the idea of
using electron scattering data to predict CC neutrino cross
sections. The model was proposed in \cite{Amaro:2004bs}, where it is
presented in detail. In summary, a phenomenological superscaling
function, extracted from QE $(e,e')$ data within a fully
relativistic framework and embodying the essential nuclear dynamics,
is multiplied by the appropriate charge-changing $N\to N$ weak
interaction cross sections in order to obtain the various response
functions contributing to the inclusive $(\nu_l,l)$ cross section,
$R_L$, $R_T$ and $R_{T^\prime}$, each response being a combination
of vector and axial-vector components.

In Figs.~\ref{fig:sigmanumu} and \ref{fig:sigmanumubar} we compare
the MiniBooNE and NOMAD QE data on $^{12}$C for $\nu_\mu$ and
$\overline\nu_\mu$ scattering, respectively, with the results of
 the RFG and SuSA models using the standard value
$M_A$=1.032 GeV/c$^2$ for the nucleon axial-vector dipole mass. The
SuSA results are represented by a narrow band, corresponding to the
uncertainty linked to the use of two fits of the phenomenological
scaling
function~\cite{Donnelly:1998xg,Donnelly:1999sw,Maieron:2001it,Amaro:2004bs}.
We observe that if the standard axial-vector mass is used the two
models underestimate the MiniBooNE data for both neutrino and
antineutrino scattering while they are both quite compatible with
the NOMAD data.

In Fig.~\ref{fig:sep} we show the breakdown of the neutrino and
anti-neutrino cross sections into individual $L$, $T$ and $T'$
contributions, with the last occurring as a positive (constructive)
term in the neutrino cross section and a negative (destructive) term
in the antineutrino cross section. Upon examining the results
displayed in Figs.~\ref{fig:sigmanumu} and \ref{fig:sigmanumubar} we
note that were this $VA$ interference to be a bit larger, for
instance via inclusion of contributions that go beyond the impulse
approximation (see above), then better agreement with the neutrino data
in the region of the MiniBooNE kinematics could be obtained, since this is 
where that term peaks, while the agreement with the antineutrino 
data would be less good.

We also show in Figs.~\ref{fig:sigmanumu} and \ref{fig:sigmanumubar} a line that
represents the cross section computed with the SuSA model but where we have excluded all
contributions coming from excitation energies below 50 MeV to assess
the importance of this region in the total cross section. 
As can be seen, this region is
quite important even for very high neutrino energies (typically
amounting to about 10\% of the total). 
As noted above, the SuSA approach was not formulated
to deal with such low-energy excitations and one might be concerned
that the present modeling is spurious for these contributions.
However, an alternative approach was taken long ago based on the
excitation of discrete ph states in the regions up through where giant
resonances dominate~\cite{JSOC72,Don85} and encouragingly the present
SuSA model and those old results essentially agree, giving us
confidence that the SuSA approach on the average does a rather good
job even at such low excitation energies. 
It should also be mentioned that  
in \cite{Ama07} the contribution of the discrete
excitations of the final nucleus $^{12}$N in CC neutrino scattering
from $^{12}$C was evaluated in a semi-relativistic shell model.  The
contribution from the discrete spectrum turned out to be below 2\%
for potential parameters fitted to reproduce the $Q$-value
of the reaction.

In Figs.~\ref{fig:sigmanumu} and \ref{fig:sigmanumubar}
the electromagnetic form factors of the nucleon entering in the vector
CC current are those of the extended Gari-Krumpelmann (GKex) model
of \cite{Lomon:2001ga,Lomon:2002jx,Crawford2010}, whose validity
extends over a wide range in the transferred momentum. In
Fig.~\ref{fig:emffnu} we compare the SuSA results obtained with
several other modern parametrizations of the form factors $G_E$ and
$G_M$~\cite{GonzalezJimenez:2011fq}. In order to appreciate the
dependence upon the vector form factors we do not show the error
band in the SuSA result and instead have inserted a sub-panel
zooming in on the region near the maximum. We observe that the
uncertainties due to the electromagnetic nucleon form factors and
the ones of the superscaling model are of the same order.
Furthermore, all of the parametrizations are essentially equivalent
for the kinematics that are relevant for these neutrino scattering
experiments.

\begin{figure}[htcb]
\includegraphics[scale=0.4, bb=20 5 722 550, clip]{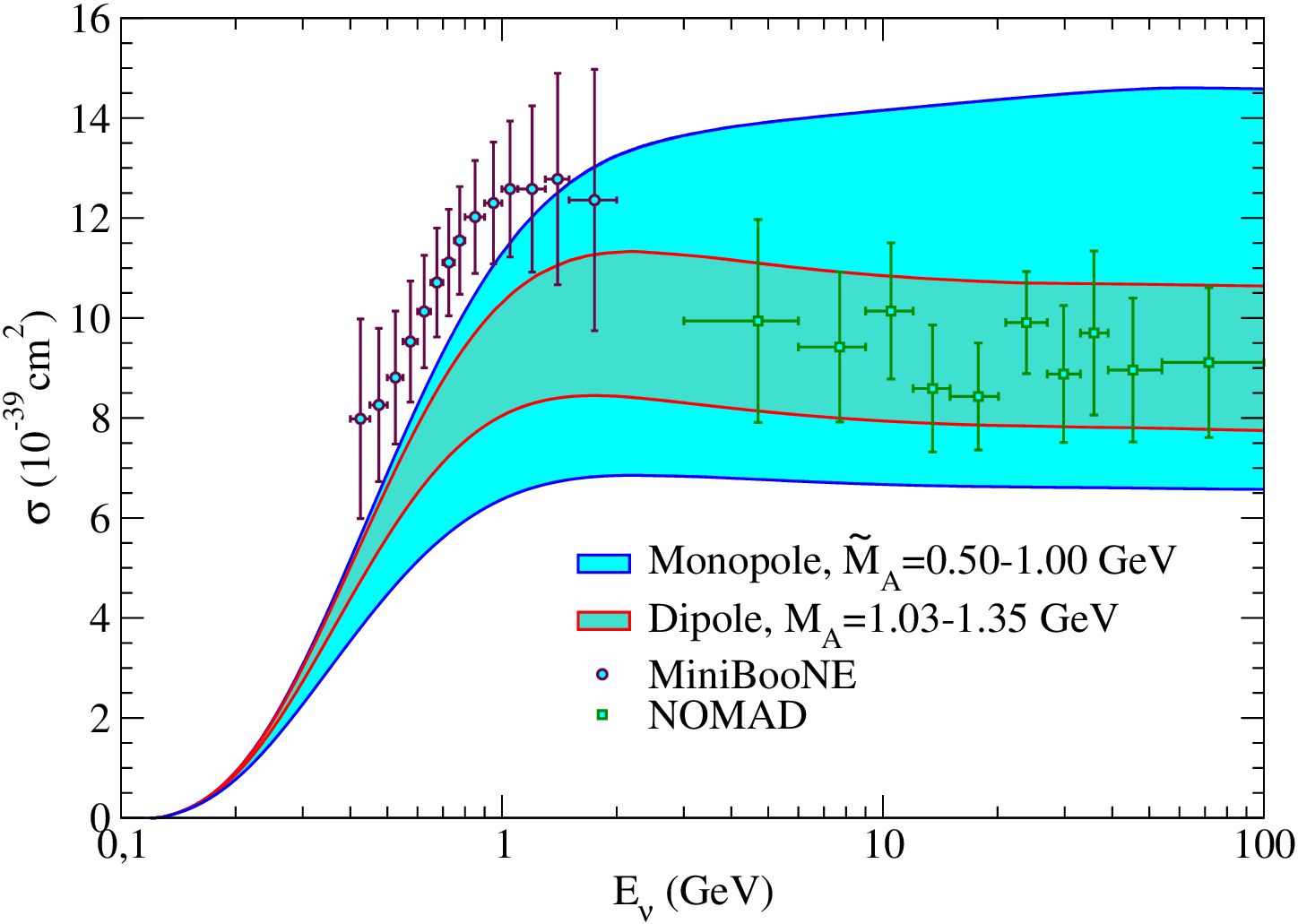} 
 \vspace{0.19cm}
\includegraphics[scale=0.4, bb=20 32 722 550, clip]{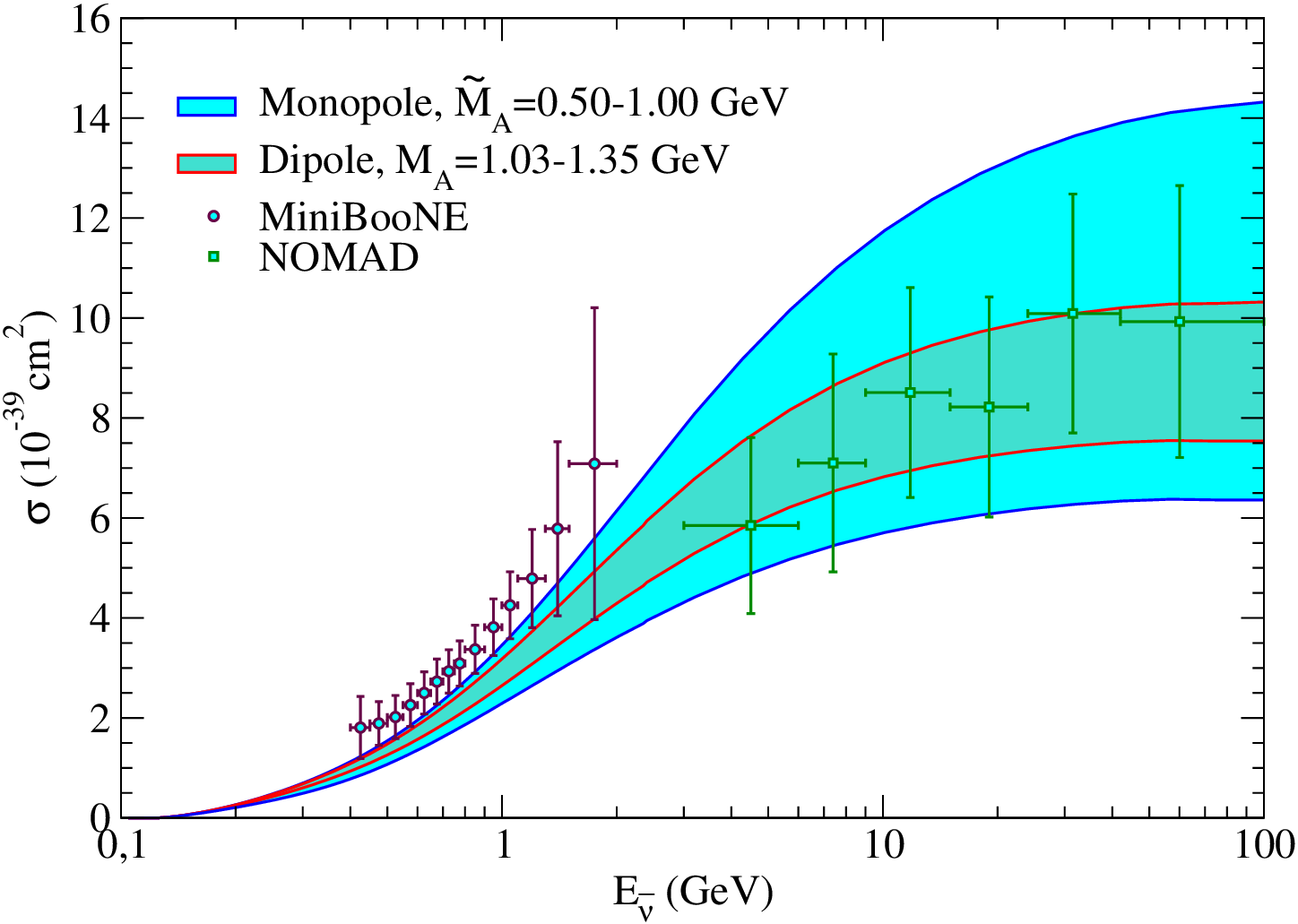} 
\caption{(Color online) CCQE $\nu_\mu-^{12}$C (upper panel) and
$\overline\nu_\mu-^{12}$C (lower panel)
 cross section per nucleon evaluated in the SuSA model for monopole (blue outer band) and
 dipole (red inner band)  parametrizations of the nucleon axial-vector form factor. A larger mass yields a higher cross section.
\label{fig:MONOP2}}
\end{figure}
\begin{figure}[htcb]
\includegraphics[scale=0.4, bb=20 32 722 550, clip]{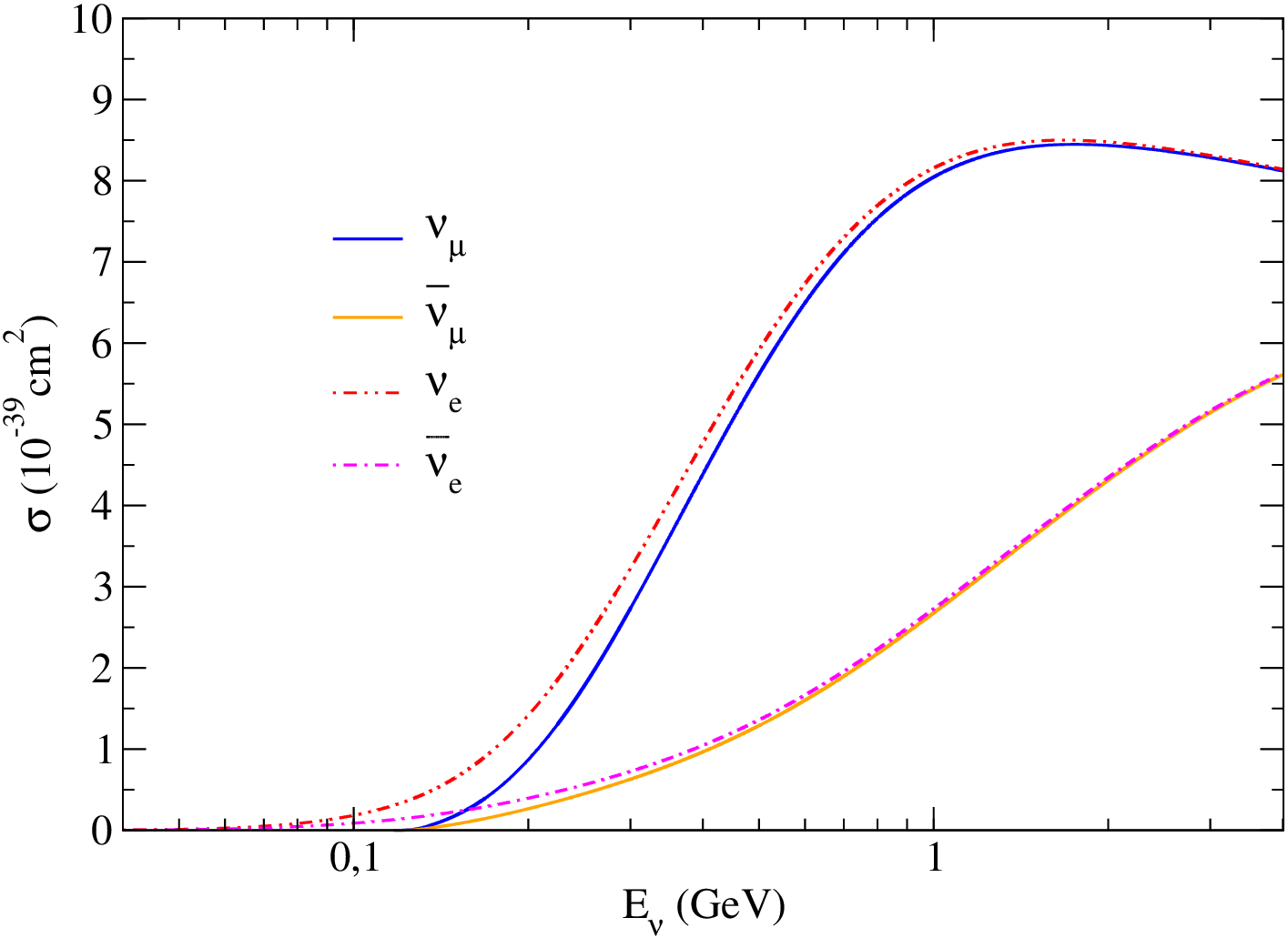} 
\caption{(Color online) SuSA predictions for muon- (solid curves)
and electron- (dotted curves) neutrino and antineutrino CCQE cross
section per nucleon on $^{12}$C. 
\label{fig:numunue}}
\end{figure}
\begin{figure}[htcb]
\begin{center}
\includegraphics[scale=0.5, bb=20 15 722 750, clip]{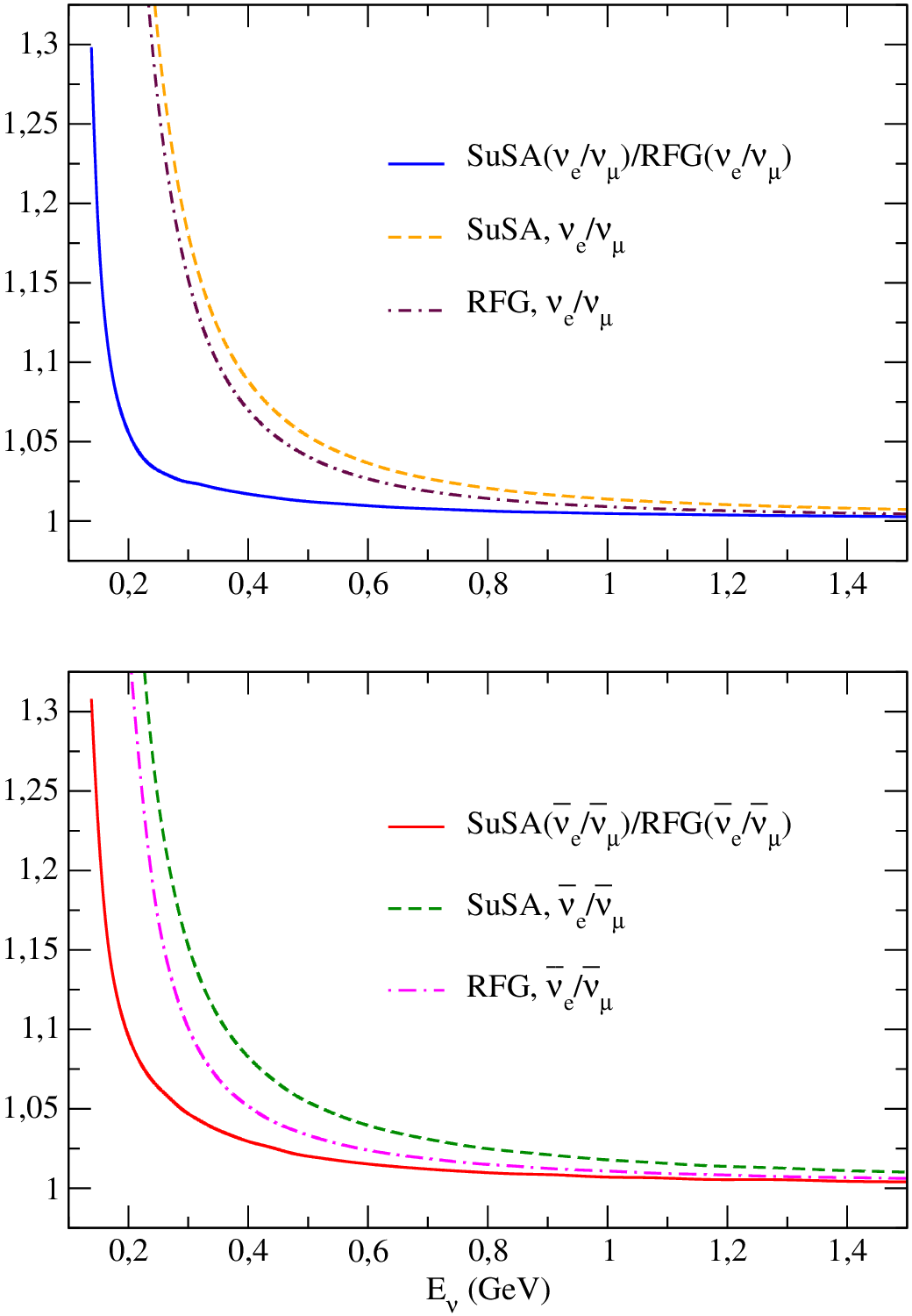}
\end{center}
\vspace{-0.35cm}
\caption{(Color online) Electron/muon neutrino (upper panel) and
antineutrino (lower panel) CCQE cross section on $^{12}$C evaluated
in the SuSA and RGF models.
\label{fig:ratios}}
\end{figure}
 Next we explore the sensitivity of the cross sections to the axial-vector form factor.
 When employing a dipole parametrization for this the ``standard'' value of the axial-vector
 dipole mass is $M_A=$1.032 GeV/c$^2$, whereas in analyzing the MiniBooNE data a large
 value of $M_A=$1.35 GeV/c$^2$ was proposed~\cite{AguilarArevalo:2010zc}. The range spanned by these
 two is shown in Fig.~\ref{fig:MONOP2} for neutrino and antineutrino
 scattering. Clearly the modified axial-vector mass produces an increase of the cross
section that allows one to fit the low-energy data in the RFG model,
although the increase is too large to explain the data at high
energy.
 Although phenomenologically successful, the dipole parametrization
 cannot be justified from a field-theoretical point of view~\cite{Masjuan:2012sk} and it is
 well-known that, for instance in vector-meson dominance (VMD) models, the fact that at moderate
 momentum transfers the EM magnetic form factors are roughly dipole-like is a conspiracy
 involving the (monopole) $\rho$ and $\omega$ poles leading to an effective dipole behaviour
 (see, for example, the discussions in~\cite{Crawford2010}).  
 Therefore, in addition to the standard
 dipole form, we also  consider a monopole form
 $G^A_M(Q^2) = [1+Q^2/{\widetilde M_A}^2]^{-1}$
 motivated by VMD-based analyses such as those in
 \cite{Lomon:2001ga,Lomon:2002jx}.  Using the monopole
 axial-vector masses $\widetilde M_A=$0.5 GeV/c$^2$ and $\widetilde M_A=$1
 GeV/c$^2$ (the range considered in~\cite{GonzalezJimenez:2011fq}) employing the SuSA model we obtain the
 band also shown in Fig.~\ref{fig:MONOP2}. Note that increasing the axial-vector
 mass produces an increase of
the cross sections with both parametrizations and a monopole
axial-vector form factor with ${\widetilde M_A}\simeq$ 1 GeV/c$^2$
leads to better agreement with both neutrino and antineutrino MiniBooNE cross sections.
On the contrary, the same model overestimates significantly the
higher-energy NOMAD data. In fact, the band width linked to the two
$\widetilde{M}_A$-values used with the monopole axial-vector form
factor is much larger than the one corresponding to the dipole
parametrization. This is in accordance with previous results shown
within the framework of parity-violating electron
scattering~\cite{GonzalezJimenez:2011fq}. We should notice that a
dipole axial-vector form factor with $M_A=1.35$ GeV/c$^2$ (in the
SuSA model) produces a cross section that is slightly lower in the
MiniBooNE energy region than that obtained using $\widetilde M_A=1$
GeV/c$^2$, but gives a ``reasonable (or a better)'' explanation of
the NOMAD data. On the other hand, $\widetilde M_A=1$ GeV/c$^2$ is
probably not a good choice because the neutrino cross section keeps
rising even at high energies. Indeed if one were to accept the
monopole parametrization and fit the NOMAD data one would find that
$\widetilde M_A=0.70 \pm 0.06$ ($0.72 \pm 0.14$) GeV/c$^2$ for neutrinos
(antineutrinos). 
Old experiments with deuterium bubble chambers also performed fits of
the data using a monopole axial form factor, obtaining 
$\widetilde M_A=0.57 \pm 0.05$~\cite{Baker:1981su}
and $\widetilde M_A=0.54 \pm 0.05$~\cite{Miller:1982qi}.
While these studies would suggest that a dipole axial-vector form factor with 
the standard value of the dipole mass
is preferred, given the modern interest in a potentially different
behaviour, especially at high momentum transfers, new studies of neutrino
disintegration of deuterium would be very valuable in clarifying this
issue. 

In Fig.~\ref{fig:numunue} we compare the $\nu_e$ ($\overline\nu_e$)
and $\nu_\mu$ ($\overline\nu_\mu$) cross sections in the SuSA model
for the kinematics relevant for the proposed facility
$\nu$STORM~\cite{Kyberd:2012iz}, which will provide high quality
electron neutrino beams in the energy range $E<4$ GeV for precise
measurements of neutrino-nucleus cross sections. In particular, this
could allow one to study the differences between muon and electron
neutrino QE cross sections. Although the hadronic interaction is the
same for $\nu_\mu$ and $\nu_e$, the different mass of the outgoing
leptons produces a different energy transfer to the nucleus for the
same incident neutrino energy.  As seen in Fig.~\ref{fig:numunue},
this results in a small shift for low neutrino energy.  For higher
energies the small differences due to the lepton mass tend to
disappear, yielding a universal curve, independent of the neutrino
flavour. This is emphasized in Fig.~\ref{fig:ratios}, where the
ratios between $\nu_e$ and $\nu_\mu$ (upper panel) and
$\overline\nu_e$ and $\overline\nu_\mu$ (lower panel) cross sections
in the SuSA and RFG models, as well as the double ratio SuSA/RFG,
are shown. We note that the different models give the same results
reaching unity for the ratio at energies above 1 GeV. For small
energies one expects that the different nuclear excitation energy
involved and the energy-dependence of the nuclear response functions
will emphasize differences between the two cross sections, or
between either of these and more sophisticated modeling of the
low-lying nuclear excitations. A precise measurement of the cross
sections in this region might therefore allow one to extract new
information concerning the electroweak nuclear matrix elements. It
should be noted that the differences between RFG and SuSA in
$\nu_e/\nu_\mu (\bar{\nu}_e/\bar{\nu}_\mu)$ (Fig.~\ref{fig:ratios})
are caused, at least partially, by the different theoretical
descriptions of the nuclear responses employed in these models,
specifically that the RFG scaling function is bounded and does not
extend to large and small values of the scaling variable.

Finally, we note that the difference between the two cases also to
some degree arises not only via the different kinematics associated
with the outgoing lepton mass, but also from Coulomb corrections,
{\it i.e.,} distortions of the final-state charged lepton wave
functions in the Coulomb field of the nucleus. These are taken into
account using the effective momentum approximation described in
\cite{Amaro:2004bs}. Their effects are found to be negligible in the
energy range considered, becoming important only at neutrino
energies below 200 MeV, where the cross sections are extremely
small. Coulomb corrections have been incorporated in all results
shown in Figs.~\ref{fig:numunue} and \ref{fig:ratios}.

\section{Conclusions}

In this paper we have shown for the first time how the superscaling
(SuSA) model behaves after being extended from intermediate to high
energies as are relevant for recent neutrino scattering experiments.
Comparisons are also made with the RFG model. Note that, although
the differences between the RFG and SuSA predictions at high energy
are small compared with the experimental error bars, the RFG fails
to reproduce the $(e,e')$ data, whereas the SuSA model agrees by
construction with the electron scattering QE cross section. Also
note that, although the region of low excitation energy plays a
significant role in the total cross section at all energies and
consequently either of these models should be viewed with caution,
since they are not well-suited to modeling the details of this
region, in fact from comparisons with discrete-state modeling the
SuSA approach does a reasonable job there when focusing on
the total cross sections.

 The SuSA model is expected to be robust enough to describe neutrino and antineutrino
 quasielastic cross sections in all of the experimentally available
 kinematic ranges, in the present study lacking only the two-body 2p2h MEC contributions
 that are expected to increase the cross sections by perhaps $10-15\%$ or
 so. 
For the reasons stated above these are not included in the
 present work but will be added later once a robust approach to
 modeling them in the low-$q$/low-$\omega$ region is in hand.

 We have presented results for cross sections for MiniBooNE and
 NOMAD conditions, and have shown predictions corresponding to the
 $\nu$STORM facility kinematics. The SuSA model has been used to
 investigate several aspects of the neutrino-nucleus interaction
 entering into the cross section, namely the impact from vector and axial-vector nucleon
 form factors and the dependence on the lepton flavour.  The
 axial-vector form factor determines the strength of the axial-vector current
 matrix elements and crucially depends on the value of the axial-vector mass
 parameter.  The dependence of the cross section upon this parameter
 is significant and yields uncertainties that are
 bigger than the other uncertainties of the model for high
 energies. Additionally, there is the issue of whether the axial-vector form factor
 should have a dipole or a monopole nature. Both types of behaviour have been explored
 in the present work. 

\section*{Acknowledgements}

This work was partially supported by 
DGI (Spain) grants FIS2011-28738-C02-01 and  FIS2011-24149, 
by the Junta de Andalucia grants FQM-170 and FQM-225, 
by the Italian INFN under contract MB31,
by the INFN-MICINN Collaboration agreement AIC-D-2011-0704,
the Spanish Consolider-Ingenio 2000 programmed CPAN, and in part (TWD) 
by US Department of Energy under cooperative agreement
DE-FC02-94ER40818. GDM acknowledges support from a fellowship from the
Fundaci\'{o}n C\'{a}mara (University of Sevilla). 
We thank R. Gonz\'{a}lez-Jim\'{e}nez for providing the various parametrizations of
the EM form factors and for useful discussions, 
and L. Alvarez-Ruso for critical reading of the manuscript and helpful
suggestions.

\end{document}